\documentclass[%
 reprint,
 superscriptaddress,
 amsmath,amssymb,
 aps,prl
]{revtex4-2}

\usepackage{graphicx}
\usepackage{dcolumn}
\usepackage{bm}
\usepackage[bookmarksnumbered, pdfpagelabels=true, plainpages=false, colorlinks=true, linkcolor=blue, citecolor=blue, urlcolor=blue]{hyperref}
\usepackage[dvipsnames]{xcolor}

\usepackage{subfloat}

\newcommand{\be}{\begin{equation}}
\newcommand{\ee}{\end{equation}}
\newcommand{\ba}{\begin{eqnarray}}
\newcommand{\ea}{\end{eqnarray}}
\newcommand{\bd}{\begin{displaymath}}
\newcommand{\ed}{\end{displaymath}}

\def\thalf{{\textstyle{\frac{1}{2}}}}

\begin{document}


\title{Scaling of the Surface Free Energy as a Probe of the QCD Critical Region}

\author{Joseph I. Kapusta}%
\email{kapusta@umn.edu}
\affiliation{School of Physics and Astronomy, University of Minnesota, Minneapolis, MN 55455, USA}

\author{Mayank Singh}
\email{mayank.singh@vanderbilt.edu}
\affiliation{Department of Physics and Astronomy, Vanderbilt University, Nashville, TN 37240, USA}

\author{Shensong Wan}%
\email{wan00589@umn.edu}
\affiliation{School of Physics and Astronomy, University of Minnesota, Minneapolis, MN 55455, USA}

\begin{abstract}
    The QCD phase diagram is expected to have a critical point that separates the crossover and first-order transition lines.
    A realistic model that incorporates phase boundary effects is essential for heavy ion simulations to isolate the experimental signatures of a critical point.
    We discuss how to construct such an equation of state, and study its critical behavior. 
    The effect of the coefficient of surface energy on the size of the critical region is investigated.  We found that with this construction and with the chosen background equation of state, the temperature must be within one percent of its critical value to observe the critical exponents.  This makes it doubtful that the critical exponents can be measured in heavy ion collisions, though it may still be feasible to observe signatures of a first-order phase transition. The method presented here is general and can be utilized with any given equation of state to test the viability of observing critical exponents in experiments.
\end{abstract}

\maketitle

\textit{Introduction--} Quantum chromodynamics (QCD) is postulated to have a first-order phase transition between the confined hadronic phase and the deconfined quark gluon plasma (QGP) phase at large baryon chemical potentials \cite{Fukushima:2010bq, Fukushima:2013rx, Fischer:2018sdj}. Lattice QCD calculations have shown that the phase change occurs via a smooth crossover at zero baryon density \cite{Aoki:2006we}. The first-order phase transition curve is expected to terminate at a critical point at some finite baryon chemical potential. 

The phase transition region of the QCD phase diagram could be accessible in intermediate-energy nuclear collision experiments and in neutron star mergers. The Relativistic Heavy-Heavy Ion Collider (RHIC) has completed the Beam Energy Scan (BES) program which gives access to the finite baryon density region of the phase diagram. The upcoming Facility for Antiproton and Ion Research (FAIR) will extend this coverage to even higher baryon densities. The hope is that one of these collision energies will bring the resulting system close enough to the critical point such that its signatures can be detected in experimental observables. If the critical point happens to be at baryon chemical potentials much larger than the coverage of collider experiments, it may be accessible in binary neutron star (BNS) mergers. Gravitational waves from a BNS merger have been observed by the LIGO and VIRGO detectors \cite{LIGOScientific:2017vwq,LIGOScientific:2017ync}. Planned new generation detectors like the Cosmic Explorer and the Einstein Telescope are expected to have a much more enhanced BNS merger detection rate. 

The search for the QCD critical point and the first-order curve in experiments relies on identifying their signatures on the observables. Realistic modeling of these experiments is necessary to identify these signatures. Models which incorporate the effects of critical dynamics and first-order phase transition need the equation of state (EOS) of QCD matter over a wide region of the phase diagram as an input. The QCD EOS at zero baryon chemical potential is known from first-principle lattice gauge theory calculations \cite{HotQCD:2014kol,Borsanyi:2013bia}. At non-zero baryon chemical potentials, the EOS has been calculated in the lattice gauge approach by Taylor expanding the QCD action around zero chemical potential \cite{Hegde:2014sta,Guenther:2017hnx,Bazavov:2017dus,Monnai:2019hkn,Noronha-Hostler:2019ayj,Borsanyi:2021sxv,Kahangirwe:2024cny}. It has also been calculated using perturbative QCD techniques \cite{Albright:2014gva,Kurkela:2009gj} and effective field theories \cite{Tews:2012fj}. Recent developments are summarized in \cite{Sorensen:2023zkk,MUSES:2023hyz,Alarcon:2025qmz}.  These approaches have different regimes of applicability, and the EOS is often interpolated between these regimes \cite{Albright:2014gva,Semposki:2024vnp,ReinkePelicer:2025vuh}. We expect the QCD critical point to be in the same universality class as the 3D Ising Model and the liquid-gas phase transition \cite{Berges:1998rc,Halasz:1998qr}. For a given critical point, the phase transition curve can be added onto the background EOS such that it results in the correct critical exponents of the universality class \cite{Guida:1996ep,Nonaka:2004pg,Parotto:2018pwx,Kapusta:2021oco,Kapusta:2022pny}.

While the EOS has information about QCD matter in a stable equilibrium phase, the interesting dynamics of the first-order phase transition happens in the metastable (via nucleation) or unstable (via spinodal decomposition) phases \cite{Randrup:2009gp}. It is essential to model the system in these phases to identify the experimental signatures of first-order phase transition. A method to extend the EOS to the metstable and the unstable regions was proposed in \cite{Kapusta:2024nii}. In this work, we use the extended EOS to relate the coefficient of surface energy ($K$) to the correlation length and the surface energy. If the surface energy is known away from the critical point, it can be used to constrain the value of $K$. The size of spinodal effects on observables is determined by the value of $K$.

In heavy-ion collisions, the cumulants of net baryon number are expected to diverge as the correlation length diverges near the critical point \cite{Stephanov:1998dy,Stephanov:1999zu,Hatta:2002sj}. This will result in non-monotonic behavior of these cumulants as a function of collision energy \cite{Stephanov:2008qz}. This has been proposed as a possible signature of the critical point. A key assumption underlying this reasoning is that the system generated at some particular collision energy reaches close enough to the critical point (i.e. in the critical region) such that it sees the critical scaling of correlation length. Our formalism can be used to explicitly calculate the correlation length as one gets closer to the critical point. We show that for our choice of the QCD EOS \cite{Kapusta:2025por}, the absolute value of $\tau \equiv (T - T_c)/T_c$ must be less than $10^{-2}$ for critical scaling to appear. Here $T$ is the temperature of the system and $T_c$ is the critical temperature. This formalism can be used to calculate how close the system needs to get to the critical point to observe its effects for any given EOS. This can also help estimate the required phase diagram resolution in collider experiments needed to identify the critical point.

\textit{QCD equation of state in metastable and unstable regions--} A covariant formulation of relativistic hydrodynamics with spinodal decomposition which can be used to describe the evolution of QCD matter was given in Ref. \cite{Kapusta:2024nii}. The same paper also provided a parameterization of the metastable and unstable regions that is consistent with mean field values of the critical exponents, specifically $\delta = 3$. For systems in the same universality class as the 3D Ising model and the liquid-gas phase transition, the more accurate value is $\delta \approx 4.79$ \cite{Guida:1996ep}.  The more general case was outlined in the appendix of Ref. \cite{Kapusta:2024nii}, specifically
\ba
P_{{\rm int}}(n)  
= P_X(T) + a_1 (n - n_G) + a_2 (n - n_G)^2 \;\;\;\;\;\nonumber\\ 
+ a_3 (n - n_G)^{\delta - 1}  
 + a_4 (n - n_G)^{\delta}
+ a_5 (n - n_G)^{\delta + 1}
\ea
and
\ba
\mu_{{\rm int}}(n) &=& \mu_X(T) + b_0 \ln(n/n_G) + b_1 (n - n_G) \nonumber\\
& & + b_2 (n - n_G)^{\delta - 1}  + b_3 (n - n_G)^{\delta}
\ea
where $P_{\rm int}$ and $\mu_{\rm int}$ are the interpolated pressure and baryon chemical potential in the coexistence region. The equilibrium pressure and chemical potential across the coexistence curve is denoted by $P_X$ and $\mu_X$. Baryon density is denoted by $n$ with $n_G$ and $n_L$ denoting its value in the hadronic gas and QGP phases, respectively. 
The parameters $a_i$ and $b_i$ are functions of temperature. At fixed $T$
\be
n \frac{\partial \mu_{\rm int}}{\partial n} = \frac{\partial P_{\rm int}}{\partial n}
\ee
This leads to
\ba
a_1 &=& b_0 + b_1 n_G \nonumber \\
a_2 &=& \thalf b_1 \nonumber \\
a_3 &=& b_2 n_G \nonumber \\
a_4 &=& \left( \frac{\delta - 1}{\delta} \right) b_2 + b_3 n_G\nonumber \\
a_5 &=& \left( \frac{\delta}{\delta + 1} \right) b_3 
\ea
There are four independent parameters to be determined at each temperature. They can be fixed in the manner described in Ref. \cite{Kapusta:2024nii} and detailed below.

\textit{Surface energy and correlation length--} Consider the behavior as the critical temperature is approached.  Since $\partial P/\partial n = \partial^2 P/\partial n^2 = 0$ at $T_c$ it must be that both $a_1$ and $a_2$ go to zero as $T \rightarrow T_c^-$. This implies that $b_0$ and $b_1$ also go to zero.  Along the critical isotherm approaching the critical point $\mu - \mu_c \sim {\rm sgn}(n - n_c) |n - n_c|^\delta$ which means that $b_2$ goes to zero as well, but $b_3$ does not.  This is consistent with $P - P_c \sim |n - n_c|^\delta$.

The density profile for a planar interface between the two phases satisfies
\be\label{eq:evolution_eq}
K \frac{d^2n}{dx^2} = b_0 \ln(n/n_G) + b_1 (n - n_G) + b_2 (n - n_G)^{\delta - 1} + b_3 (n - n_G)^{\delta}
\ee
while the surface energy $\sigma$ is
\be\label{eq:surface_energy}
\sigma = K \int_{-\infty}^{\infty} dx \left(\frac{dn}{dx}\right)^2.
\ee
Here $K$ is the coefficient of surface energy. In general the surface profile can only be found numerically.  However, if the system is so close to $T_c$ that $\Delta n = n_L - n_G \ll n_c$ then the equation can be linearized in each of the two phases.  In the gas phase we write $n - n_G = \varepsilon$ with $0 < \varepsilon \ll n_c$.  The linearized equation is
\be
\frac{d^2\varepsilon}{dx^2} = \frac{b_0 + n_G b_1}{K n_G} \varepsilon \equiv \frac{\varepsilon}{\xi_G^2} 
\ee
The quantity $\xi_G^2$ should be positive so that the solutions are real exponentials rather than sines and cosines.  In the liquid phase we write $n - n_G = \Delta n - \varepsilon$ with $0 < \varepsilon \ll n_c$.  Then the linearized equation is
\be
\frac{d^2\varepsilon}{dx^2} = - \frac{b_0 + n_L b_1}{K n_L} \varepsilon \equiv \frac{\varepsilon}{\xi_L^2} 
\ee
which follows after using the condition
\ba
& b_0 \ln(n_L/n_G) + b_1 (n_L - n_G) + b_2 (n_L - n_G)^{\delta - 1}\nonumber\\ & + b_3 (n_L - n_G)^{\delta} = 0
\ea
to the appropriate order.  Note the change in sign between $b_0 + n_G b_1$ and 
$b_0 + n_L b_1 = b_0 + n_G b_1 + \Delta n \, b_1$.

Supposing that the phase for $x < 0$ is the liquid and the phase for $x > 0$ is the gas then the approximate solutions are
\ba\label{eq:solution_n}
n(x < 0) &=& n_L - a_L \, {\rm e}^{x/\xi_L} \nonumber \\
n(x > 0) &=& n_G + a_G \, {\rm e}^{-x/\xi_G}
\ea
where both $a_L$ and $a_G$ are positive.  Continuity of the density and its first derivative result in
\ba
a_L + a_G = \Delta n \nonumber \\
\frac{a_L}{\xi_L} = \frac{a_G}{\xi_G}
\ea
The surface free energy is therefore approximately
\be
\sigma = \frac{K}{2} \left( \frac{a_L^2}{\xi_L} + \frac{a_G^2}{\xi_G} \right)
\ee

From the above analysis one might expect that $a_L \sim a_G \sim \Delta n \sim (-\tau)^{\beta}$ and 
$\xi_L \sim \xi_G \sim (-\tau)^{-\nu}$.  Therefore the expectation is that 
$\sigma \sim (-\tau)^{2\beta + \nu}$.  The definition of the critical exponent $\bar{\mu}$ for the surface energy is 
$\sigma \sim (-\tau)^{\bar{\mu}}$ \footnote{This critical exponent is usually called $\mu$ but to avoid confusion with a chemical potential it is sometimes called $\bar{\mu}$, which is what we do here.}.  The estimate from this analysis is that $\bar{\mu} = 2\beta + \nu$.  It also follows from this analysis that $\sigma$ scales as $\sqrt{K}$.  The numerical value used in Ref. \cite{Kapusta:2024nii} was 
$K = 5\times10^{-5}$ MeV$^{-4} \approx 1.4959\times10^{7}$ MeV fm$^5$.

A well-known textbook calculation in mean field approximation for any model in the same universality class yields $\bar{\mu} = 3/2$.  An $\epsilon$ expansion yields $\bar{\mu} = 3/2 - \epsilon/4 \rightarrow 1.25$ \cite{Brezin:1984zz}.  
Taking $\alpha = 0.1101$ and $\beta = 0.3264$ from \cite{El-Showk:2014dwa, Simmons-Duffin:2015qma} as used in \cite{Kapusta:2022pny} and in this paper, and using the scaling relations $\alpha + 2\beta + \gamma = 2$ and
$\gamma = \beta (\delta -1)$, 
we get $\gamma = 1.2371$ and $\delta = 4.7901$.
Measured values of $\bar{\mu}$ for atomic and molecular systems in this universality class, which involve challenging experiments, are in the range from 1.23 to 1.28 \cite{Pelissetto:2000ek}.

Note that the exponents of $b_0$ and $b_1$ are very close to $\gamma - \beta = 0.9107$, and the exponent of $b_2$ is very close to $\beta$.  Therefore we expand these functions near the critical point as
\ba
b_0(\tau) &=& b_{00}(-\tau)^{\gamma - \beta} + b_{01}(-\tau)^{\gamma} + \cdot\cdot\cdot \nonumber \\
b_1(\tau) &=& b_{10}(-\tau)^{\gamma - \beta} + b_{11}(-\tau)^{\gamma} + \cdot\cdot\cdot
\ea
where the $b_{ij}$ are constants.  Continuity of the first derivative at phase coexistence is
\be
\frac{\partial P}{\partial n}(n_G) = \frac{n_G}{\chi_G} = a_1
\ee
and
\ba
\frac{\partial P}{\partial n}(n_L) &=& \frac{n_L}{\chi_L} = a_1 + 2 a_2 \Delta n
+ (\delta -1) a_3 (\Delta n)^{\delta-2} \nonumber\\
& & + \delta a_4 (\Delta n)^{\delta-1} + (\delta +1) a_5 (\Delta n)^{\delta}
\ea
As $\tau \rightarrow 0^-$ the densities are symmetrically located at
\ba
n_G &=& n_c - \thalf \Delta n \nonumber \\
n_L &=& n_c + \thalf \Delta n
\ea
with
\be
\Delta n = n_0 (-\tau)^\beta
\ee
where $n_0$ is a constant.  Then $n_G/\chi_G \sim n_L/\chi_L \sim (-\tau)^\gamma$ requires that $b_{00} +n_c b_{10} = 0$.  Corrections to scaling of $b_0$ and $b_1$ are of relative order $(-\tau)^\beta$ which can account for the small discrepancies extracted numerically in the critical exponents.  Actually it can be seen from the above equations that
\ba\label{eq:analytic_xi}
\frac{1}{\xi_G^2} &=& \frac{a_1}{K n_G} = \frac{1}{K \chi_G} \nonumber \\
\frac{1}{\xi_L^2} &=& - \frac{a_1 + b_1 \Delta n}{K n_L} = \frac{1}{K \chi_L}
\ea
in the limit that $\tau \rightarrow 0^-$.  Hence in this EOS $\nu = \gamma/2 = 0.6186$.  
Fisk and Widom \cite{FiskWidom1969} calculated that $\bar{\mu} = \gamma - \nu + 2 \beta$.  Our estimate of $\bar{\mu}$ above is in agreement with this when using $\nu = \gamma/2$.  Numerically then $\bar{\mu} = 1.2714$.

Along the coexistence curve the susceptibilities on the liquid and gas sides are equal. Therefore 
$\chi_L(T) = \chi_G(T) \equiv \chi_X(T)$ is a function of $T$ only.  If we change variable from $n$ to $\mu$ we have
\be
\frac{dn}{dx} = \chi_X(T) \frac{d\mu}{dx}
\ee
and so
\be
K \chi_X \frac{d^2\mu}{dx^2} = \mu
\ee
showing that
\be\label{eq:correlation_length}
\xi^2 = K \chi_X
\ee
is the same on both sides of the coexistence curve.

Experimental searches for the QCD critical point typically depend on the critical scaling behavior that arises as the system approaches it. A key question is whether theory can indicate how close the system must be to the critical point to reliably observe scaling behavior. What are the associated length and energy scales? The answer depends on the value of $K$, which is a fundamental property of QCD matter. We consider $K$ as a parameter and select three distinct values that span three orders of magnitude: $10^5$, $10^6$, and $10^7$ MeV fm$^5$.

\begin{figure}
    \centering
    \includegraphics[width=0.45\textwidth]{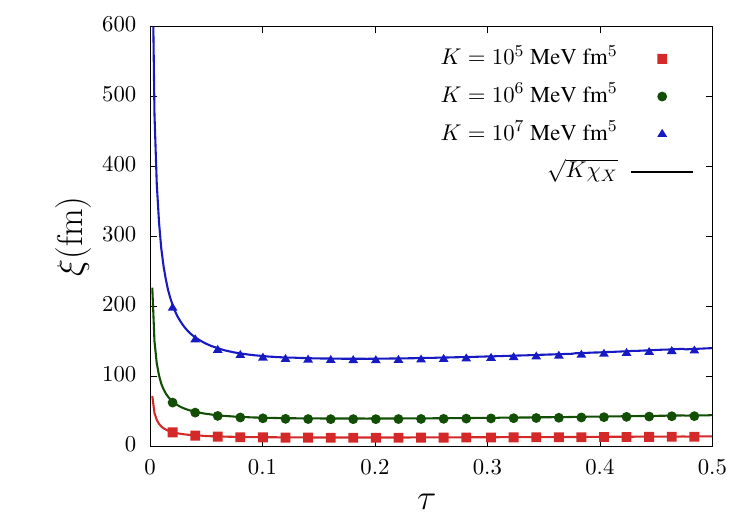}
    \includegraphics[width=0.45\textwidth]{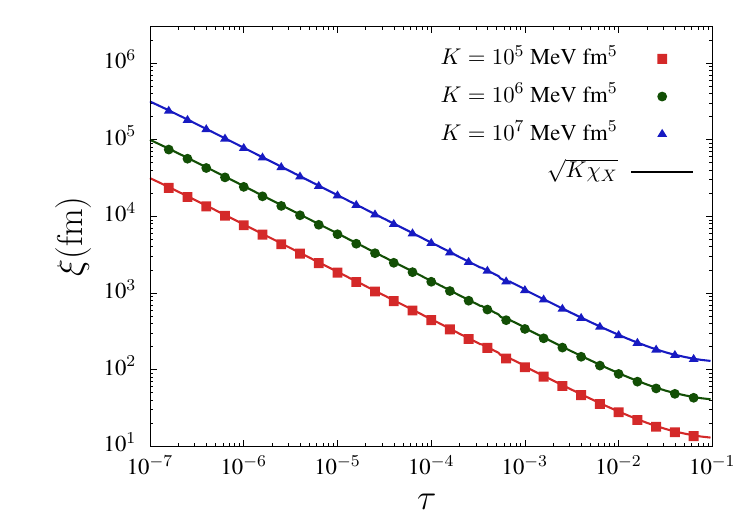}
    \caption{Correlation length compared with its analytic expression from Eq. (\ref{eq:correlation_length}) for different values of the temperature in linear (top) and log (bottom) scales.  The solid curves represent the expression from Eq. (\ref{eq:correlation_length}). We recover the correct critical exponent for $\tau \lessapprox 10^{-2}$.}
    \label{fig:correlation}
\end{figure}

Given an equation of state, Eq. (\ref{eq:evolution_eq}) can be solved numerically. We use the EOS formulated in Ref. \cite{Kapusta:2025por} with the critical point at $T_c = 120$ MeV and $\mu_c = 670$ MeV, and the phase boundary is determined by condition A given in the reference. As a second order ordinary differential equation, it needs two boundary conditions. We start with $n(x=0) = (n_G + n_L)/2$. For the second condition, we do a search for values of $dn/dx$ at $x=0$ such that $n \rightarrow n_L^-$ as $x \rightarrow +\infty$ and $n \rightarrow n_G^+$ as $x \rightarrow -\infty$. We consider the solution 
\be\label{eq:solution_evolution_eq}
n(x) = \thalf (n_L + n_G) + \thalf \Delta n \tanh (x/2\xi),
\ee
which can be fit to the numerical solution of Eq. (\ref{eq:evolution_eq}) to obtain the correlation length $\xi$. Figure \ref{fig:correlation} shows the correlation lengths obtained from numerical fits compared to the analytic expression in Eq. (\ref{eq:correlation_length}). There is good agreement between analytic and numerical results.

\begin{figure}
    \centering
    \includegraphics[width=0.45\textwidth]{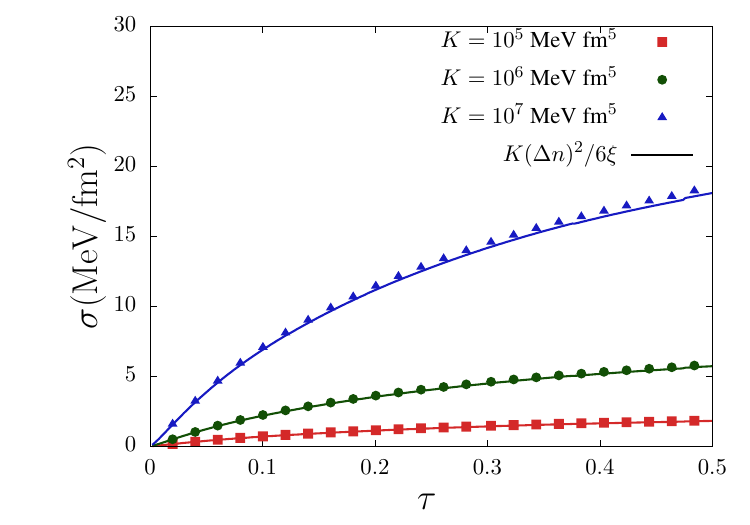}
    \includegraphics[width=0.45\textwidth]{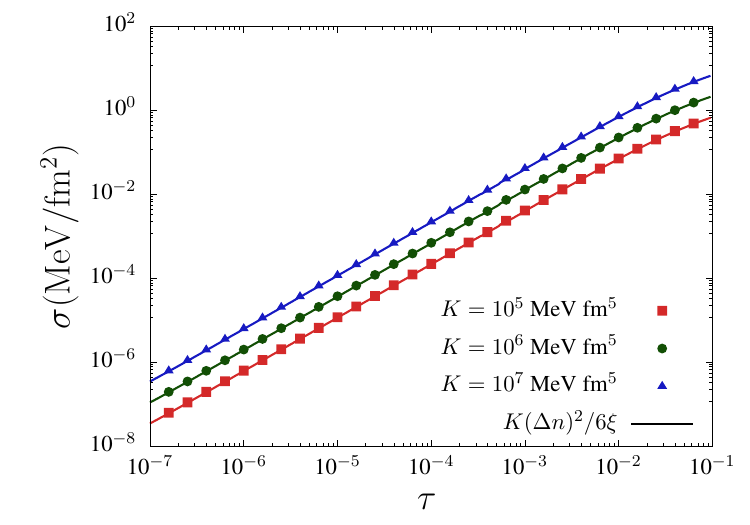}
    \caption{Surface free energy compared with its analytic expression from Eq. (\ref{eq:surface_energy_analytic}) for different values of the temperature in linear (top) and log (bottom) scales. The solid curves represent the expression from Eq. (\ref{eq:surface_energy_analytic}). We recover the correct critical exponent for $\tau \lessapprox 10^{-2}$. }
    \label{fig:sigma}
\end{figure}

When the correlation length is plotted on a log scale for small $\tau$, we see the critical scaling appears for $\tau \lessapprox 10^{-2}$. So, at least for this EOS, the system will need to be very close to the critical point to observe its effects. For $K = 10^5$ MeV fm$^5$, the correlation length is on the order of tens of fm at $\tau = 10^{-2}$ when critical scaling starts appearing. For larger values of $K$, it is much bigger. This is larger than the typical system size in heavy-ion collisions. Hydrodynamical models suggest that the portion of the system that gets this close to the critical point is even smaller \cite{De:2022yxq}. It appears that it will be very difficult for a heavy-ion collision system to feel the effects of critical scaling. However, the effect could be very important for BNS mergers.

The choice of $K$ can be motivated by the value of surface energy. Using the parameterization in Eq. (\ref{eq:solution_evolution_eq}) with Eq. (\ref{eq:surface_energy}), 
the surface free energy is
\be\label{eq:surface_energy_analytic}
\sigma = \frac{K (\Delta n)^2}{6 \xi}
\ee
In the scaling region $\sigma \propto \sqrt{K}$ and $\sigma \propto (-\tau)^{\bar{\mu}}$ with $\bar{\mu} = 2\beta + \nu$.

The surface energy and its comparison with its analytic expression in Eq. (\ref{eq:surface_energy_analytic}) is plotted in Fig. \ref{fig:sigma}.  For the largest value of $K = 10^7$ MeV fm$^5$, at $T = 0.5T_c$, the surface energy is 26 MeV/fm$^2$. The surface energy at the same temperature for $K = 10^5$ MeV fm$^5$ is about 2.6 MeV/fm$^2$. Again, the critical scaling appears for $\tau \lessapprox 10^{-2}$. So, at least for this EOS, the system would need to be very close to the critical point to observe its effects.

\begin{figure}
    \centering

    \includegraphics[width=0.45\textwidth]{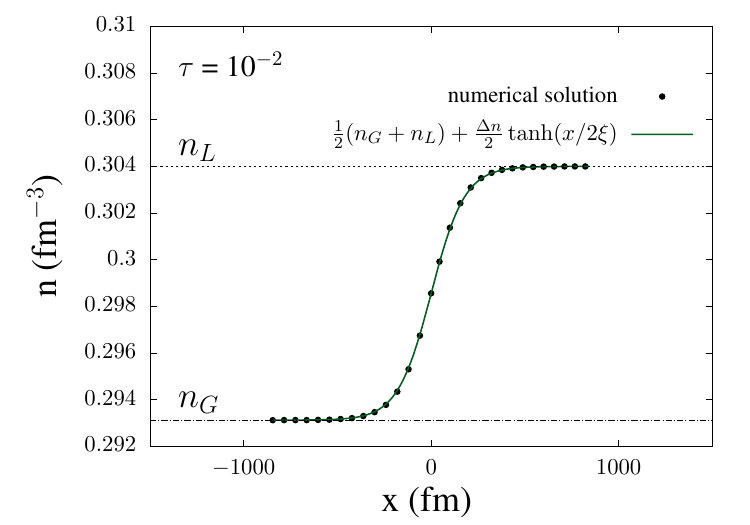}
    \includegraphics[width=0.45\textwidth]{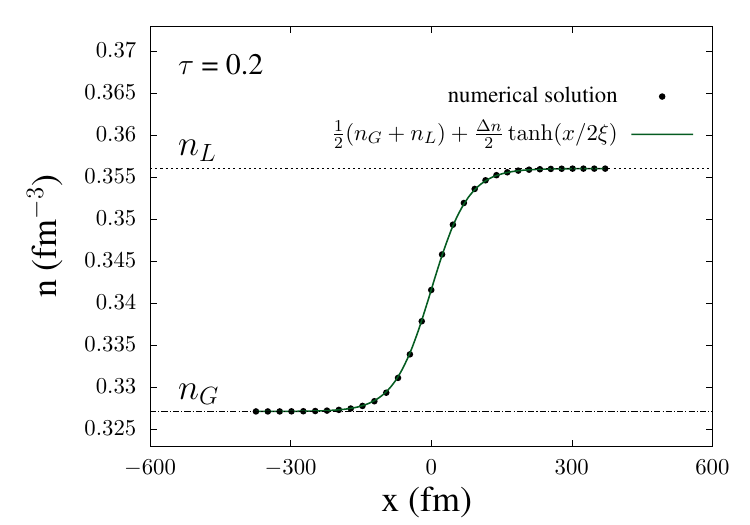}
    \caption{Baryon density as a function of position inside the surface obtained from solving Eq. (\ref{eq:evolution_eq}). The solution is matched to Eq. (\ref{eq:solution_evolution_eq}) for different values of the temperature.}
    \label{fig:n_evolution}
\end{figure}

We can also define the surface thickness. The common 90-10 definition of the surface thickness $t$ is when the density rises 10\% above $n_G$ until it reaches 10\% below $n_L$.  Then
\ba
n\left(\thalf t\right) &=& {\textstyle{\frac{1}{10}}} (9 n_L + n_G) \nonumber \\
n\left(-\thalf t\right) &=& {\textstyle{\frac{1}{10}}} (9 n_G + n_L)
\ea
leads to 
\be\label{eq:surface_thickness}
t = 4 \xi \tanh^{-1}(4/5) = 4.394 \, \xi
\ee
Figure \ref{fig:n_evolution} shows good agreement between the numerical solution and Eq. (\ref{eq:solution_evolution_eq}) for a wide range of $\tau$. As expected, the surface thickness increases as the temperature approaches its critical value.

The energy and length scales depend on the specific choice of the EOS and on the value of $K$. The interpolation method and the subsequent calculations of surface energy and correlation lengths in this paper can be used to estimate the size of the critical region for any proposed equation of state. This will help assess the feasibility of measuring observables that indicate critical scaling in a given experiment.

\section*{Acknowledgements} This work was supported by the U.S. DOE Grants
No. DE-FG02-87ER40328 (J.I.K. and S.W.) and No. DE-SC-0024347 (M.S.).

\bibliography{ref}{}

\end{document}